\documentstyle[12pt,epsf]{article}  
\newcommand{\op}{{\overline\phi}}

\newcommand{\be}{\begin{equation}}
\newcommand{\ee}{\end{equation}}
\newcommand{\ben}{\begin{eqnarray}\displaystyle}
\newcommand{\een}{\end{eqnarray}}
\newcommand{\refb}[1]{(\ref{#1})}

\newcommand{\sectiono}[1]{\section{#1}\setcounter{equation}{0}}

\begin{document}

{}~ \hfill\vbox{\hbox{hep-th/0008227}
\hbox{CTP-MIT-3018}}\break

\vskip 3.5cm

\centerline{\large \bf A Solvable Toy Model for Tachyon Condensation}
\vspace*{1.5ex}

\centerline{\large\bf in String Field
Theory}

\vspace*{10.0ex}

\centerline{\large \rm Barton Zwiebach
\footnote{E-mail: zwiebach@mitlns.mit.edu}}

\vspace*{1.5ex}

\centerline{\large \it Center for Theoretical Physics}
\centerline{\large \it Massachusetts Institute of Technology}
\centerline{\large \it  Cambridge, MA 02139, USA}

\vspace*{4.5ex}
\medskip
\centerline {\bf Abstract}

\bigskip
The lump solution of $\phi^3$ field theory
provides a toy model for unstable D-branes
of bosonic string theory. The field theory
living on this lump is itself a cubic 
field theory involving a tachyon, two
additional scalar fields, and a scalar field continuum.
Its action can be written explicitly because
the fluctuation spectrum of the lump
turns out to be  governed by a solvable 
Schroedinger equation; the $\ell=3$ case
of a series of reflectionless potentials. 
We study the multiscalar tachyon potential both exactly
and in the level expansion, obtaining insight
into issues of convergence, branches of the solution
space, and the mechanism
for removal of states after condensation. In particular
we find an interpretation for the puzzling finite
domain of definition of string field marginal
parameters.

\vfill \eject
\baselineskip=17pt

\tableofcontents  

\sectiono{Introduction and summary} \label{s1}

It has recently been realized that Sen's conjectures
on tachyon condensation and D-brane annihilation 
\cite{9902105,9805019,9805170,9808141}
can be studied quite effectively using string field 
theory (SFT) 
\cite{WITTENBSFT,KS,9912249,0001201,0002237,0002117,
0003031,0005036,0008053,0008101,
0001084,0002211,0003220,0007153,0008033,0008127}.
With strong evidence now available to the physical correctness of the
conjectures, we also have the opportunity to use tachyon dynamics to
refine our understanding of string field theory. While the
level expansion studies of the bosonic D-brane tachyon 
potential find with high accuracy the requisite critical point,
an analytic solution, showing
conclusively the existence of this critical point is
still lacking.   Such solution for the
tachyon condensate would surely teach us a lot
about the nature of the string field equations and would help
finding many other interesting solutions. 

One way to make progress with this question is to develop
techniques to deal with the full string field equations.
These techniques would presumably use the universality
of the tachyon potential \cite{9911116} thus requiring methods
where various Virasoro algebras play a central role, as
in \cite{0006240}. Another line of investigation is to use toy models
and/or simplified models of the tachyon condensation phenomenon.
Particularly interesting have been a study in the framework
of $p$-adic open string theory \cite{0003278} and papers where
non-commutativity is added via a magnetic field on the 
D-brane \cite{0005006,0005031,0006071,0007226,0008013,0008064}. 

In this
paper we develop a toy model which captures a different aspect of the
problem: the decay of a lump as seen by the field theory of the
lump itself. Bosonic D-branes are lump
solutions of string field theory. The field theory on
a D-brane itself is indeed
a string field theory, and contains a multiscalar
tachyon potential. In finding the critical point of this
string field potential we are indeed
exploring the decay of a lump from the viewpoint of the
field theory living on the world-volume of the lump. 

We begin with the simple $\phi^3$ theory, which is the
truncation of open SFT to the tachyon
field only.  As analyzed numerically in \cite{0002117} this field
theory has a codimension one lump solution that provides a first
approximation to the codimension one brane of string theory.
Additionally, being an unstable lump, the field theory living on the
lump has again a tachyon with an estimated $m^2 \approx -1.3$
\cite{0002117}. In fact,  the exact lump profile in
$\phi^3$ theory is readily written in terms of hyperbolic
functions. More surprising, however, is that the Schroedinger
type equation for fluctuations that determines the spectrum
of the field theory on the lump is exactly solvable.
The associated Schroedinger potential is actually the $l=3$
case of the infinite series of exactly solvable reflectionless
potentials:\footnote{I am grateful
to Jeffrey Goldstone for providing this identification, and 
teaching me how to solve elegantly for the spectrum of such
hamiltonians. For a pedagogical  
review on these and other solvable Hamiltonians with references
to the early literature see \cite{9405029}. Applications of 
reflectionless systems to fermions can be found in \cite{9408120}.} $U_\ell (x)  
= - \ell ( \ell + 1) \hbox{sech}^2 x$. 
In fact, it is known that the fluctuation 
spectrum of the sine-Gordon soliton is governed by the $\ell=1$
potential and the the spectrum of the $\phi^4$ kink solution
is governed by the $\ell=2$ potential \cite{christ}. In both  cases
the field theory on the soliton has no tachyon since the soliton
is stable. If one tries to associate the $\ell=3$ potential to 
a stable soliton the result is a very strange field theory potential
with branch cuts \cite{christ,GJ}. Apparently $U_3$ was not
know to be relevant to the unstable lump of the very simple
$\phi^3$ theory.

This solvability allows us to find the complete field
content of the $\phi^3$ lump field theory. This includes
a tachyon with $m^2 = -5/4$, a massless scalar, a massive
scalar with $m^2 = 3/4$, and  a continuum spectrum of
scalar fields with $m^2 \geq 1$. Given that we know the analytic 
expressions for the fluctuation eigenfunctions, the exact
cubic multiscalar field theory living on the lump can
be calculated exactly.   

We then make the following claim: {\it The multiscalar
tachyon potential of the unstable lump field theory must have 
a critical point with a negative energy density equal
to the energy density of the lump itself}. Indeed, the field
theory of the lump describes the fluctuations of the
lump, and one possible fluctuation of the unstable lump
is the no-lump configuration.
Being a critical point in field space with
zero energy density, this configuration must appear as a lower
energy state of the lump field theory. This remark 
shows that the critical point of the tachyon field theory
of the lump is easily found. Let $\overline\phi(x)$ represent
the profile of the lump itself and let $\phi_0$ denote the 
expectation value of the field at the zero energy vacuum,
with $\lim_{x\to \pm \infty} \overline\phi(x) = \phi_0$.
In addition, let the fluctuation
fields around the lump 
be written as $\sum_n \phi_n \psi_n(x)$ where the $\psi_n(x)$'s
are the eigenfunctions of the Schroedinger problem and
the $\phi_n$'s are the fields living on the lump. 
Then the critical point of the tachyon potential is
defined by the equation: 
$\phi_0 -\overline\phi(x) = \sum_n \phi_n \psi_n(x)$ which
fixes the $\phi_n$'s. Namely,
{\it the expectation values for the lump fields at
the tachyonic vacuum are given as the expansion coefficients
of minus the lump profile in terms of the eigenfunctions 
of the lump fluctuation equations.} 

Equipped with both a level expansion approach to
the multiscalar potential and the exact solution for the
tachyon condensate, we explore the convergence of the
level expansion in this model and find the large level
behavior of the expectation values of fields representing
the condensate. We are also able to track explicitly the
masses of the tachyon and other fields as the lump goes
through the annihilation process. The tachyon and other
fields on the lump flow and join eventually the continuum
spectrum of states associated to the massive field
defined on the locally stable vacuum of the $\phi^3$ model.
These results are consistent with the discussion of ref.~\cite{9805150} 
of the annihilation of a kink and an anti-kink of $\phi^4$ field theory.

Our study of the toy model gives a plausible 
resolution to the
puzzles found by Sen and the author \cite{0007153}
concerning the definition of marginal fields in
string field theory. While conformal field theory
marginal parameters are naturally defined over
infinite ranges,  it was found 
that the effective potential for the string field marginal
parameter fails to exist beyond a critical value. This 
appeared to mean that either SFT compresses
the CFT moduli into a compact domain, or that
SFT does not cover all of CFT moduli space.
By studying the toy model we are led to propose that the map
from the CFT marginal moduli to SFT marginal moduli 
is actually two to one; as the CFT parameter 
parameter grows from zero to infinity the SFT parameter 
grows from zero to a maximum value and then decreases
back to zero! On the way back, a different solution branch for
high level fields must be chosen. This behavior is borne
out by the analysis of how the massless state in the
lump, whose wavefunction is the derivative of the
profile, implements translations of the lump. Indeed,
even for large translations the expectation value for
this massless state is always bounded. 

This paper is organized as follows. The lump solution and the
spectrum is presented in section 2. The multiscalar
potential for the fields living on the brane is computed
in section 3. A detailed analysis of the tachyon condensation
and the removal of the lump states by the condensation
process is given in section 4. The puzzle of marginal
parameters is analyzed in section 5, along with a 
suggestion for a solution branch describing large marginal
deformations. This paper concludes in section 6 with a discussion
of the results.

\sectiono{The ${\bf\phi^3}$ Lump and its fluctuation spectrum}\label{s2}

In this section we will consider $\phi^3$ scalar field
theory and find the exact lump profile as well as
the exact lump energy.
We then turn to the fluctuations around the lump solution,
identifying the resulting Schroedinger like equation.
This Schroedinger equation turns out to be solvable, both
for the discrete and continuum spectrum. We give the 
explicit energy eigenfunctions thus identifying the spectrum
of the field theory living on the lump worldvolume.

\subsection{The lump solution}
\label{s21}

\begin{figure}[!ht]
\leavevmode
\begin{center}
\epsfbox{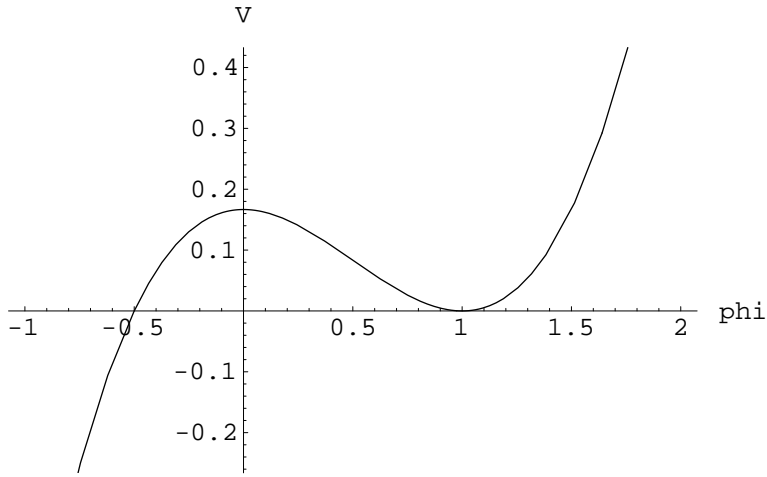}
\end{center}
\caption[]{\small The cubic potential $V(\phi)$. The tachyonic
vacuum is at $\phi=0$ and the locally stable vacuum
is at $\phi=1$.} \label{f1}
\end{figure}

We begin with the $\phi^3$ scalar field theory in $p+2$ space-time dimensions
with action
\be
\label{action}
S = \int dt d^{p}y dx \Bigl[ \, {1\over 2} \Bigl( {\partial\phi\over
\partial t}\Bigr)^2 - {1\over 2} \nabla_{\vec y} \phi \cdot \nabla_{\vec
y}
\phi  -   {1\over 2} \Bigl( {\partial\phi\over \partial x}\Bigr)^2 -
V(\phi) 
\Bigr]\,,
\ee
where we have separated out the $x$ coordinate, to be used
to produce the lump solution. The resulting lump will therefore
represent a $p$ brane, with $\vec y$ denoting the $p$ spatial
coordinates of the brane worldvolume. In the action above, the potential
will be taken to be:
\be
\label{potential}
V(\phi) = {1\over 3 \phi_*} (\phi- \phi_*)^2 \,\, (\phi + {\phi_*\over 2})
= - {1\over 2} \,\phi^2  + {1\over 3\phi_*}\, \phi^3 + {\phi_*^2\over 6}\, .
\ee
This is a generic $\phi^3$ potential, it has a local maximum and
a local minimum rather than 
an inflection point. The local maximum is at $\phi=0$, where
we have a tachyon of $m^2 = -1$ (this particular value can be
thought as a choice of units).
The local minimum is at $\phi= \phi_*$, where we have a scalar 
particle of $m^2 = +1$ (this equality of squared masses up to a 
sign, is generic). The open string field theory action, truncated 
to the tachyon and with $\alpha'=1$, is given as $S/g_0^2$, where
$g_0$ is the open string coupling constant, and $S$ is the above
action with  $p=24$ and $\phi_* = 1/K^3$ with $K=3\sqrt{3}/4$. 
By rescaling the field variable
as $\phi\to \phi_* \phi$, the action and potential can be simply
written as 
\ben
\label{finpot}
S &=& (\phi_*)^2\int d^{p+1}\hskip-2pt y\, dx \Bigl[ \, -{1\over 2} 
\partial_\mu \phi \,\partial^\mu \phi
-   {1\over 2} \Bigl( {\partial\phi\over \partial x}\Bigr)^2 - V(\phi) 
\Bigr]\,,\cr\cr V(\phi) &=& {1\over 3} \,\,(\phi- 1)^2 \,\, \bigl(\phi +
{1\over 2}\,\bigr) =\,\, {1\over 6} - {1\over 2} \,\phi^2 + {1\over 3}\,
\phi^3\,
\,,
\een
where $\phi_*$ appears now as an overall multiplicative 
constant in the action, and therefore, it does not appear
in the equations of motion. All energies will be measured
in units of $(\phi_*)^2$, which, for the purposes of the present
paper will be set to unity.  In the above equation
we have defined $y^\mu = (t,\vec y)$ with metric 
$(-, +,+, \cdots +)$. The potential
is shown in Fig.~\ref{f1}.

\begin{figure}[!ht]
\leavevmode
\begin{center}
\epsfbox{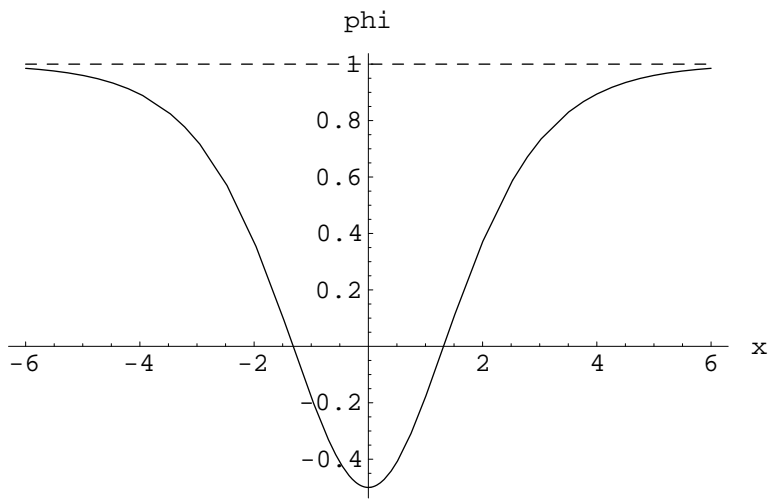}
\end{center}
\caption[]{\small The profile of the lump solution $\op (x)$. As 
$x\to \pm \infty$, $\op \to +1$, which is the expectation value
for the locally stable vacuum.} \label{f2}
\end{figure}

In the lump solution $\overline \phi(x)$,
as $x\to \pm \infty$ we must have  
$\op (x) \to 1$, which corresponds to the
locally stable vacuum. Given the familiar
mechanical analogy of motion on the potential
$(-V)$, we will find
that $\op$ will vary from $+1$ down to $(-1/2)$ 
and back to $+1$ as $x$ goes from minus to plus
infinity. For symmetry, we take $\op (x=0) = -1/2$. 
The equation one must solve is:
\be
\label{etosolve}
{d^2 \op\over dx^2 } - V'(\op) = 0 \quad \to \quad
{1\over 2} \Bigl( {d\op\over dx} \Bigr)^2 = V(\op) 
\quad\to \quad x = \int_{-1/2}^\op {d\phi'\over \sqrt{2 V(\phi')}}\,.
\ee
With the potential in \refb{finpot}, the above integral
is elementary\footnote{This is because two of the three possible
zeroes in $V(\phi)$ coincide. If the lump was placed on 
a circle of finite radius, the relevant potential would have
three different zeroes, and the integral would be expressed
in terms of complete elliptic functions.} and
gives the profile: 
\be
\label{lumpprofile}
\op (x) = 1 - {3\over 2} \, \hbox{sech}^2 \bigl( {x\over 2} \bigr)\,,
\ee
whose plot is shown in Fig.~\ref{f2}. We record, in passing, that
\be
\label{secder}
V'' (\op ) = 1 - 3\, \hbox{sech}^2 \bigl( {x\over 2} \bigr)\,.
\ee

\medskip
We now expand the action \refb{finpot} around the lump solution
by letting $\phi \to \op + \phi$, where, with a little abuse of
notation we now use $\phi$ to denote the fluctuation field
around the lump. We find (recall we have set $\phi_* =1$)
\ben
\label{pertaction}
S = \int d^{p+1}\hskip-2pt y\, dx 
\Bigl[ -   {1\over 2} \Bigl(
{d\op\over dx}\Bigr)^2 - V(\op) 
\, -{1\over 2} \partial_\mu \phi\,\partial^\mu
\phi  -   {1\over 2} \, \phi \Bigl(
 - {\partial^2 \phi\over \partial x^2} + V''(\op ) \, \phi\Bigr) - {1\over
3}\,
\phi^3\,
\Bigr]\,.
\een
The first two terms of this action give (minus) the energy density of the
$p$-brane defined by the lump solution. Indeed, the total energy $E$
is given by
\be
E = \int d^{p}\hskip-1pt \vec y\, dx 
\Bigl[  {1\over 2} \Bigl(
{d\op\over dx}\Bigr)^2 + V(\op)\Bigl]  = (\hbox{Vol}_y) \int dx \Bigl(
{d\op\over dx}\Bigr)^2 \,,
\ee
and performing the last integral one finds
\be
\label{energylump}
T_p = {E\over (\hbox{Vol}_y)} =\,  {6\over 5} \,,
\ee
for the tension of the $p$-brane. Given that the original
$\phi^3$ theory around the unstable vacuum $\phi=0$
is supposed to represent the
space-filling ($p+1$)-brane, we have $T_{p+1} = V(\phi=0) = 1/6$.
Therefore
\be
{1\over 2\pi}\,\, {T_p\over T_{p+1}} = {18\over 5 \pi} \simeq 1.146\,,
\ee
a ratio that in string theory takes the value of unity\footnote{In
\cite{0002117} the
exact value of the string theory brane tension $T_p$ is compared 
to the value $T_p^\ell$ obtained with the lump arising from the
tachyon truncation of the string action. For this ratio 
we obtain the exact value $T^\ell_p /T_p = {8192 \pi\over 32805} \simeq
0.7845$, in agreement with the numerical estimate in\cite{0002117}.}.

\medskip
The fluctuations on the brane require an analysis of the
quadratic terms in \refb{pertaction}.  For this we consider
the Schroedinger type eigenvalue equation
\be\label{schro}
 - {d^2 \psi_n\over d x^2} + V''(\op (x) ) \, \psi_n (x)= 
M_n^2 \, \psi_n(x)\,.
\ee
The relevance of this equation is that 
expanding the fluctuation field
$\phi (y, x)$ as 
\be
\label{fexp}
\phi (y, x)= \sum_n \xi_n(y) \psi_n(x)\,,
\ee 
and substituting back in \refb{pertaction}, the 
fields $\xi_n(y)$ living on the lump
would be seen to have mass squared $M_n^2$ (this analysis will
be done in section \ref{s3}).   Using the
expression for $V''(\op )$ in \refb{secder} we have
\be\label{schr}
 - {d^2 \psi_n\over d x^2} + \Bigl(1 - 3\, 
\hbox{sech}^2 \bigl( {x\over 2} \bigr) 
\, \Bigr)\psi_n (x)= M_n^2 \, \psi_n(x)\,.
\ee
Letting $x \equiv 2u$ this equation is written as
\be\label{mainsch}
 - {d^2 \psi_n\over d u^2} + \Bigl(9 - 12\, \hbox{sech}^2 u
\, \Bigr)\psi_n (u)= (4M_n^2+ 5) \, \psi_n(u)\,,\quad x \equiv 2u\,.
\ee
This equation is the  $\ell=3$ case of the following
Schroedinger problem
\be\label{schrodinger}
 - {d^2 \psi_n\over d u^2} + \Bigl(\ell^2 - \ell (\ell + 1)\, 
\hbox{sech}^2 u
\, \Bigr)\psi_n (u)= E_n (\ell) \, \psi_n(u)\,.
\ee
We now turn to the solution of this eigenvalue equation.

\subsection{$\ell=3$ wavefunctions and
lump spectrum}
\label{s22}

The eigenvalue equation \refb{schrodinger} is readily
solvable in terms of special functions.  Indeed, it is
a textbook case (see, for example \cite{LL}, section 23, problem 5)
where the wavefunctions can be written in terms of associated Legendre
polynomials when $\ell$ is an integer, and in terms of hypergeometric
functions when it is not. The solution to be presented here (which I
learned from J. Goldstone)  uses methods explained in 
generality  in \cite{9405029}. One introduces a set of operators
\be
\label{oscillators}
a_\ell = \ell \tanh u + {d\over du} \,, \quad
a_\ell^\dagger = \ell \tanh x - {d\over du}\,,
\ee
and verifies that
\be
\label{hamilt}
H_\ell \equiv    - {d^2 \over d u^2} +
\Bigl(\ell^2 -
\ell (\ell + 1)\, 
\hbox{sech}^2 u \Bigr)\, = a_\ell^\dagger  a_\ell \,. 
\ee
In addition, one readily confirms that:
\be
H_{\ell -1} = a_\ell  a_\ell^\dagger - (2\ell-1)\,. 
\ee
The equation we must solve is:
\be
H_\ell \, \psi^{(\ell)}_n (u ) = E^{(\ell)}_n \psi^{(\ell)}_n (u )\,,
\ee
for the various eigenfunctions labelled by $n$, with $n=0$ denoting
the ground state. The
ground state wavefunction
$\psi_0^{(\ell)} (u)$ is found from the condition
\be
a_\ell\, \psi_0^{(\ell)} (u) = 0 \quad \to \quad \psi_0^{(\ell)} (u)
= N_0(\ell) \,\hbox{sech}^\ell u\,.
\ee
The normalization of this wavefunction follows from the
relations
\be
I_\ell \equiv \int_{-\infty}^{\infty} du \, \hbox{sech}^\ell u, \quad
I_{\ell + 2} = {\ell\over \ell+1} I_\ell\,, \quad I_1= \pi, \quad I_2 =
2\,.
\ee
Thus, for example, for our $\ell=3$ case of interest, we have
\be
\psi_0^{(3)} (u) 
= \sqrt{15\over 16}  \,\,\hbox{sech}^3 u \,, \quad  E_0^{(3)} =0.
\ee 
To find the other wavefunctions, we use another feature of
the hamiltonians $H_\ell$:  each  $H_{\ell -1}$ eigenfunction
provides an $H_\ell$ eigenfunction.
Indeed, one readily verifies that
\be
\label{implic}
H_{\ell -1} \psi = E^{(\ell-1)} \psi \quad \to \quad
H_{\ell } (a_\ell^\dagger \psi) = (E^{(\ell-1)}+ 2\ell -1)
(a_\ell^\dagger \psi)  \equiv E^{(\ell)} (a_\ell^\dagger \psi) \,.
\ee
Applying this for $\ell=3$, the ground state
$\psi_0^{(2)} \sim \hbox{sech}^2 u$ of $H_2$ is used to find
\be
\psi_1^{(3)} \sim a_3^\dagger\, \psi_0^{(2)} \sim \tanh u \,\hbox{sech}^2
u \,,
\ee
which upon normalization yields
\be
\label{firstex}
\psi_1^{(3)}(u) = \sqrt{15\over 4} \,\, \tanh u \,\hbox{sech}^2
u \,, \quad E_1^{(3)} = 5\,.
\ee
Similarly, the ground state $\psi_0^{(1)} \sim \hbox{sech} u$ of $H_1$
can be used to find, after two steps:
\be
\label{secondex}
\psi_2^{(3)}(u) = \sqrt{3\over 16} \,\,\bigl( 
5\,\hbox{sech}^3 u - 4\,\hbox{sech} u \bigr) \,, \quad E_2^{(3)} = 8\,.
\ee
The three states above are the only bound states.
Their wave-functions are orthonormal,
$\psi^{(3)}_0$ and
$\psi^{(3)}_2$ are $u$ even and  $\psi^{(3)}_1$ is $u$ odd.
In addition to these bound states, we have a continuum of 
$\delta$-function normalizable eigenfunctions. These can
be found analogously starting with the $\delta$-function 
normalizable continuum $e^{iku}$ of 
$H_0 = - {d\over du^2} $. We therefore have
\be
\label{fcont}
\psi^{(3)}_k(u) = {1\over N(k)} \,a_3^\dagger a_2^\dagger a_1^\dagger
\,e^{iku}\,,
\quad  N(k) = \sqrt{(1+ k^2)(4+ k^2)(9+ k^2) } \,, \quad E_k = 9 + k^2\,,
\ee
satisfying the orthonormality condition
\be
\int du  (\psi^{(3)}_{k'}(u))^* \psi^{(3)}_{k}(u)  = 2\pi \delta (k-k')\,.
\ee
More explicitly, we have
\be
\label{gcont}
\psi^{(3)}_k(u) = {1\over N(k)} e^{iku} \Bigl[ \tanh u \Bigl( 6 - 6 k^2 -
15 \hbox{sech}^2 u \Bigr) + ik \Bigl( k^2 - 11 + 15 \hbox{sech}^2
u\Bigr) \Bigr]\,.  
\ee
The state at $k=0$ is $u$ odd, and while it is only $\delta$-function
normalizable, the wavefunction approaches a constant at infinity.
It should be thought as a bound-state at threshold. It will play no
special role in our discussion.

\medskip
Having found all the relevant wavefunctions, we now summarize them
and use them to provide the explicit solutions of our original
equation \refb{mainsch}.
For this we use $x$ rather than $u$, and give the
corresponding mass squared values
$M_n^2$ via the relation $E_n = 4 M_n^2 + 5$.
We also drop the $\ell=3$ index, and
use different symbols to denote $x$-even as opposed to $x$-odd
wavefunctions:
\ben
\label{summdisc}
\xi_0 (x) 
&=& \sqrt{15\over 32}  \,\,\hbox{sech}^3 \bigl({x\over 2}\bigr) \,, \quad 
M^2 = - {5\over 4}\, , \cr\cr
\eta_0 (x) &=& \sqrt{15\over 8} \,\, \tanh ({x\over 2}\bigr)
\,\hbox{sech}^2 ({x\over 2}\bigr) \,, \quad M^2 = 0\, , \cr\cr
\xi_1(x) &=& \sqrt{3\over 32} \,\,\Bigl( 
5\,\hbox{sech}^3 ({x\over 2}\bigr) - 4\,\hbox{sech} ({x\over 2}\Bigr)
\bigr) \,, \quad M^2 = {3\over 4}\,.
\een
We have one tachyon, one massless
scalar and one massive scalar. While the 
original tachyon in the $\phi^3$ theory had $m^2=-1$,
the tachyon living on the lump worldvolume has 
a larger mass $m^2= -5/4$. In string theory both 
tachyons have the same mass.
For the continuum, from \refb{fcont} and
\refb{gcont}, we introduce:
\be
\Xi_k (x)={e^{ikx/2}\over \sqrt{2} N(k)}  \Bigl[ \tanh
({x\over 2}\bigr)
\Bigl( 6 - 6 k^2 - 15 \hbox{sech}^2 ({x\over 2}\bigr) \Bigr) + ik \Bigl(
k^2 - 11 + 15
\hbox{sech}^2 ({x\over 2}\bigr)\Bigr) \Bigr]\,,  
\ee
with mass squared given by 
\be
\label{mcont}
M_k^2 = 1 + {k^2\over 4}\,.
\ee
The wavefunctions $\Xi_k(x)$ satisfy
the reality and orthogonality properties
\be
\label{reXi}   
\Xi_k^*(x) = \Xi_{-k}(x)\,,\quad \int dx\,\, \Xi_{k'}^*(x)\,\Xi_k (x) = 2\pi
\delta(k-k')\,.
\ee 
We introduce real and imaginary parts
\be
\label{continuum}
\Xi_k (x) \equiv \eta_k(x) + i \xi_k (x)\,,
\qquad  
\ee
where
\be
 \eta_{-k}=\eta_k \,, \quad  \xi_{-k}=-\xi_k  \,,\quad\hbox{and}\quad
\eta_k(-x) = -\eta_{k}(x) \,, \quad \xi_k(-x) =  \xi_k(x)\,.
\ee 
The wavefunctions $\xi_k$ and $\eta_k$  are associated with a continuum of
scalar fields with mass squared greater than or equal to one.

\sectiono{The action living on the ${\bf\phi^3}$ lump}
\label{s3}

In order to find the complete field theory living on
the worldvolume of the lump we must expand the fluctuating
field $\phi$ in terms of lump fields and lump wavefunctions,
and substitute back into the action given in \refb{pertaction}.

We begin by expanding the fluctuating field (as suggested
in \refb{fexp})  using the notation introduced above.  This
gives
\be
\label{fieldexp}
\phi(y, x) = \phi_0 (y) \, \xi_0 (x) + \psi_0 (y) \, \eta_0 (x)
+ \phi_1(y) \, \xi_1 (x) 
 + \int_{-\infty}^{\infty} dk \, \Phi_k (y) \, \Xi_k (x)\,.
\ee
Here $\phi_0$ is the tachyon, $\psi_0$ is the massless scalar
and $\phi_1$ is the massive scalar, all living on the lump
worldvolume. 
Reality of $\phi(x,y)$, on account of \refb{reXi}, requires that
\be
\Phi_k^*(y) = \Phi_{-k}(y)\,\, \to\,\, \Phi_k(y) \equiv \psi_k(y) + i
\phi_k(y)\,,\quad\hbox{with}\quad \phi_{-k} = -\phi_k, \quad \psi_{-k}=
\psi_k\,. 
\ee

\medskip
We will focus our attention on the exact potential for the fields
living on the lump. For this, it suffices to consider the part
of the action in \refb{pertaction} that includes the last two
terms:
\be
\label{pertpot}
V = \int d^{p+1}\hskip-1pt y \Bigl[\,\int dx \Bigl\{
  {1\over 2} \, \phi \Bigl(
 - {d^2 \phi\over d x^2} + V''(\op ) \, \phi\Bigr) + {1\over 3}\, \phi^3\,
\Bigr\} \Bigr]\,.
\ee
In order to find the potential for the standard fields (as opposed
to the continuum fields) we must simply substitute 
$\phi = \phi_0  \, \xi_0 (x) + \psi_0 \, \eta_0 (x)
+ \phi_1 \, \xi_1 (x)$ into the above, and perform
the integral over $x$.  
The resulting potential takes the form $V= \int d^{p+1}\hskip-1pt y\,
V_{dis}$, with
\ben
\label{discpot}
V_{dis} &=& - {5\over 8} \, \phi_0^2  + {3\over 8} \, \phi_1^2 \cr\cr
&& + {175\over 8192} \sqrt{15\over 2}\, \pi\, \phi_0^3 
+ {225\over 8192} \sqrt{3\over 2} \,\pi \,\phi_0^2 \, \phi_1 
+ {129\over 8192} \sqrt{15\over 2} \,\pi \,\phi_0 \, \phi_1^2 \cr\cr
&&-  {201\over 8192} \sqrt{3\over 2} \,\pi \, \phi_1^3
+ \Bigl( {75\over 2048} \sqrt{15\over 2}\, \pi\,  \phi_0 
-{105\over 2048} \sqrt{3\over 2}\, \pi\,  \phi_1 \Bigr) \psi_0^2\,. 
\een
Note the masslessness of $\psi_0$, expected as it represents
the translation mode of the lump. 

\medskip
For the continuum states we can also determine the potential. 
We focus on the quadratic terms only. Making use of
\refb{mcont}, \refb{reXi}, and \refb{continuum}, 
and writing $V= \int d^{p+1}\hskip-1pt y\,
V_{cont}^{(2)}$,  we find
\ben
V^{(2)}_{cont} &=& {1\over 2} \int \, dx
\int_{-\infty}^{\infty} dk' \,
\Phi_{k'} (y) \, \Xi_{-k'}^* (x) \Bigl[ -{d^2\over dx^2} + V''(\op) \Bigr]
\int_{-\infty}^{\infty} dk \, \Phi_k (y) \, \Xi_k (x)\, \cr\cr
&=& \pi \int_{-\infty}^{\infty} dk  \,
\Phi_{-k} (y) \,\Bigl(1 + {k^2\over 4} \Bigr)
 \Phi_k (y)  \cr\cr
&=& 2\pi \int_{0}^{\infty} dk  \,\Bigl[
\phi_{k}^2  
+  \psi_{k}^2 \,\Bigr] \,\Bigl(1 + {k^2\over 4} \Bigr)  \,.
\een

\medskip
We can calculate all other interaction terms involving
both discrete and continuum fields, as well as continuum
fields only.  Since our analysis will not make use of all
of such terms and the expressions are considerably lengthy
we will not attempt to record their explicit forms.

\sectiono{Analysis of tachyon condensation}
\label{s4}

In this section we begin by studying the decay
of the lump with a level expansion analysis of
the multiscalar tachyon potential. This is, to date,
the only tool available in SFT. We then turn to an
exact analysis, based on the fact that the vacuum
state represents a well defined fluctuation of the
lump profile. Finally, we discuss the flow of the
masses of various states, including the tachyon,
as the lump decays away.

\subsection{Analysis in the level expansion}

As is customary we will assign level
zero to the tachyon field $\phi_0$. Our first approximation to the 
condensation problem is to work at level zero, where
we have the mass term and the cubic interaction 
of the tachyon $\phi_0$ (see \refb{discpot}):
\be
V^{(0)}(\phi_0) = - {5\over 8} \, \phi_0^2 
 + {175\over 8192} \sqrt{15\over 2}\, \pi\, \phi_0^3 \,.
\ee
The nontrivial critical point is at
\be
\label{zlt}
\overline\phi_0 = {2048\over 105 \pi} \,\sqrt{2\over 15} \simeq 2.26705\,,
\ee
giving us
\be
V^{(0)} (\op_0)= -\, {1048576\over 99225\, \pi^2} \simeq  -1.07073\,.
\ee
The absolute value of this result is our first
approximation to the tension of the lump, given
in \refb{energylump}. Thus forming the ratio, whose
exact value should be unity, we find
\be
{|V^{(0)} (\op_0)|\over T_p} = {524288\over 59535 \pi^2} \simeq
0.89227\,.
\ee
which at about 90\%, is surprisingly close to the expected
answer (in bosonic string theory this first
approximation gives about 70\% of the vacuum energy \cite{9912249}). 

To continue, recall that twist odd states
play no role in tachyon condensation to the stable vacuum
in open string theory. For completely analogous reason,
states that arise from wavefunctions that are odd under
$x\to -x$, such as $\psi_0$, will not acquire expectation
values. So we can restrict ourselves to $\phi_0$ and $\phi_1$.
Since the mass squared of $\phi_1$ is $3/4$ and that of $\phi_0$
is $(-5/4)$, the field $\phi_1$ must be assigned level two ($=3/4 -(-5/4)$). 

We can therefore work out the (2,4) approximation (fields up to level two,
and interactions up to level 4). This includes all terms involving $\phi_0$
and $\phi_1$ except for the term cubic in $\phi_1$. This time we find
\be
\label{24exp}
\hbox{Level (2,4)}: \qquad  \op_0 \simeq 2.41575, \quad \op_1 \simeq
-0.43908\,,
\ee
giving us
\be
\quad V^{(2,4)} (\op) \simeq -1.1917\, \quad \to \quad  {|V^{(2,4)}
(\op)|\over T_p}
\simeq 0.9931\,. 
\quad 
\ee
Indeed, this gives an extremely close value for the energy density
of the lump. We now try the level (2,6) approximation, by including
the $\phi_1^3$ term. This gives
\be
\label{26exp}
\hbox{Level (2,6)}:  \qquad \op_0 \simeq 2.405, \quad \op_1 \simeq
-0.403234\,,
\ee
resulting in
\be
\quad V^{(2,6)} (\op) \simeq -1.1185\,\quad \to \quad {|V^{(2,4)}
(\op)|\over T_p}
\simeq 0.9872\,. 
\quad 
\ee
It may seem surprising that the error in the energy density has
increased. This is presumably because once we are using level
six interactions the continuum spectrum (whose fields start at
level 9/4) could have played a significant role. At any rate
we will see later on that the addition of the cubic term $\phi_1^3$
has brought the expectation value of $\phi_1$ significantly closer 
to its true value.

\begin{figure}[!ht]
\leavevmode
\begin{center}
\epsfbox{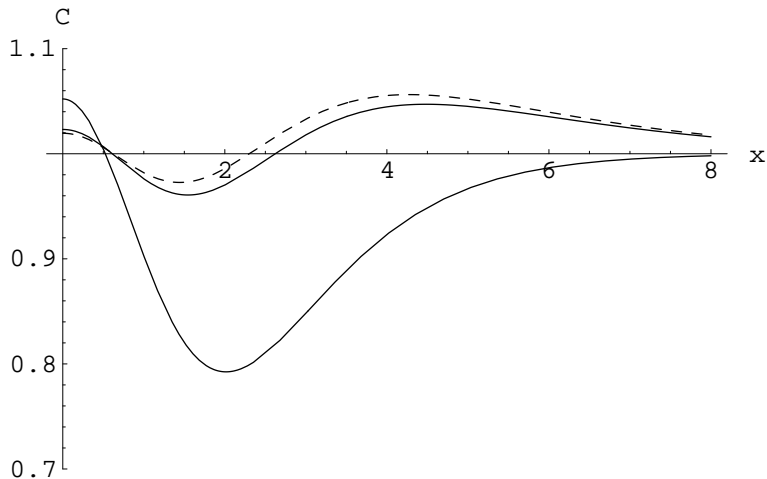}
\end{center}
\caption[]{\small Plot of the function $C(x)$ defined
in \refb{errf}. At level zero we have the curve that
extends down to about 0.8 (recall that the profile $\op$
went all the way down to -0.5). The level (2,4) curve
is dashed, and the level (2,6) curve is continuous
and close to the (2,4) curve.} \label{f3}
\end{figure}

\medskip
It is instructive to see how the above solutions are
attempting to reconstruct the vacuum solution $\phi(x) =1$.
We can consider the function $C(x)$ defined as
\be
\label{errf}
C(x) = \op (x) + \op_0 \, \xi_0(x)  + \op_1 \, \xi_1(x) \,.
\ee
This function is computed by adding to the profile $\op (x)$
the appropriately weighted wavefunctions of the first two
states. As explained in the introduction, this sum evaluated
at the exact expectation values, and including additionaly the continuum
states should reproduce the  vacuum configuration:
$\op (x\to \infty) = 1$. Since we have only included
the effect of the discrete fields, and their approximate
expectation values $C(x)$ is only expected to be roughly
equal to one.  The plot of 
$C(x)$ for the various approximations we have taken is shown in
Fig.~\ref{f3}. While the original profile $\op (x)$, shown in
Fig.~\ref{f2} extends from $-0.5$ to $1.0$, we see that at level
(2,4) (or (2,6)) $C(x)$ has already been flattened around one,
extending less than 5\% up or down.  

\subsection{Exact analysis of the condensation}

The exact condensate is found from the condition that the
fluctuation representing the condensate added to the profile
$\op(x)$ must give the vacuum configuration $\phi(x)=1$.  Therefore
we have  the equation
\be
\label{exactcond}
1-\op(x)  ={3\over 2}\, \hbox{sech}^2 \bigl( {x\over 2} \bigr)  = \op_0  \,
\xi_0 (x) +
\op_1 \, \xi_1 (x) 
 - 2 \int_{0}^{\infty} dk \, \op_k \, \xi_k (x)\, ,
\ee
where $\op_0, \op_1,$ and $\op_k$ are the desired expectation values
for the lump fields representing the stable vacuum as seen from the
lump.  Using the orthogonality of the associated $\xi$
eigenfunctions we find:
\ben
\label{vacvalues}
\op_0 &=&  {9\over 32} \, \sqrt{15\over 2} \pi \simeq 2.41976 \,,\cr\cr
\op_1 &=&  -{3\over 8} \, \sqrt{3\over 32} \pi \simeq -0.3607 \,,\cr\cr
\op_k &=&  {3\over 16\sqrt{2}}\,\, {k^2\over \hbox{sinh} (k\pi /2)}
%\Biggl( {4+k^2\over (1+k^2) (9+ k^2) } \Biggr)^{1/2}\,.
\,\sqrt{ {4+k^2\over (1+k^2) (9+ k^2) }} \,\, .
\een
Note that for the continuum states \refb{mcont} the level $L$ is given
by $L(\op_k) = 1 + {k^2\over 4} + {5\over 4}$ (see \refb{mcont}). Taking
the large $k$ limit of the amplitude $\op_k$ we find
\be
\op_k \sim  \sqrt{L} \exp ( - \pi \sqrt{L} )\,.
\ee
Since all  fields have properly normalized kinetic
terms, this gives an idea of how fast the expectation
values of fields decay as we go up in the level expansion.
There is still no analogous result in the string field theory--in this
case the number of states themselves increase as the level
increases. In the present model we have two 
states at each continuum level,
one arising from an even wavefunction, and one arising
from an odd wavefunction.

\medskip
We can do  a  consistency check
by using the following simple result: given a
multiscalar potential consisting of quadratic and cubic terms only,
the quadratic terms evaluated at the solution give a value
three times larger than the value of the potential at the
solution. For our case, we must therefore have that the value
$(-6/5)$ of the potential at the minimum can be written as
\be
-{6\over 5} = {1\over 3}\, \Bigl[ - {5\over 8} \,{\op_0}^2 
+ {3\over 8}\,  {\op_1}^2 
+ 2\pi \int_0^\infty dk\,\phi_k^2 \Bigl( 1 + {k^2\over 4} \Bigr) \Bigr]\,,
\ee 
where the terms inside brackets in the right hand side are
the quadratic terms in the potential evaluated at the solution.
Using the values quoted in \refb{vacvalues} we must have
\be
\label{spectral}
-{6\over 5} =-\, {999 \pi^2\over 8192} + {3\pi\over 1024} 
\int_0^\infty dk\,{k^4 (4+k^2)^2 \over (1+k^2) (9+ k^2) \sinh^2 (k\pi/2)}\,.
\ee 
While the definite integral in the right hand side can surely
be done analytically, the correctness of the above equation has
been verified numerically to high accuracy. Equation \refb{spectral}
can be viewed as a kind of spectral decomposition of the lump energy.
The continuum lump spectrum is seen to carry about 0.298\% of the
lump energy, a small part indeed.

\subsection{Tracking the flow of the states}

In this subsection we will study the fate of the
fields living on the lump as the tachyon condenses.
We will focus on the fields $\phi_0$ and $\phi_1$,
namely, the tachyon and the massive scalar. 
We will imagine giving expectation values to the tachyon $\phi_0$
that vary from zero to $\op_0$.  
As in the discussion of \cite{9805150}, 
we imagine adding a source term that cancels the
tadpole arising because we are not at stationary
points of the potential. For every value of $\phi_0$
we solve for $\phi_1$ using its equation of motion.  
Some intuition to the generic behavior of the flow follows from
the discussion in \cite{9805150}. One expects both
fields to eventually merge into the continuum and there
will be a particular configuration along the flow
where the tachyon will have zero mass.

We consider first directly expanding around
the tachyonic vacuum.  To first approximation we can
simply work to level zero, where $\op_0 = 2.267$ (see \refb{zlt}).
Expanding the level zero potential around this expectation value 
by  letting
$\phi_0 \to 2.267 + \phi_0$ we find that
\be
V^{(0)} \simeq -1.071 + 0.625 \phi_0^2 + {\cal O}(\phi_0^3)\,,
\ee
which implies a $m^2 \simeq 1.25$ for the tachyon after
condensation.  

\begin{figure}[!ht]
\leavevmode
\begin{center}
\epsfbox{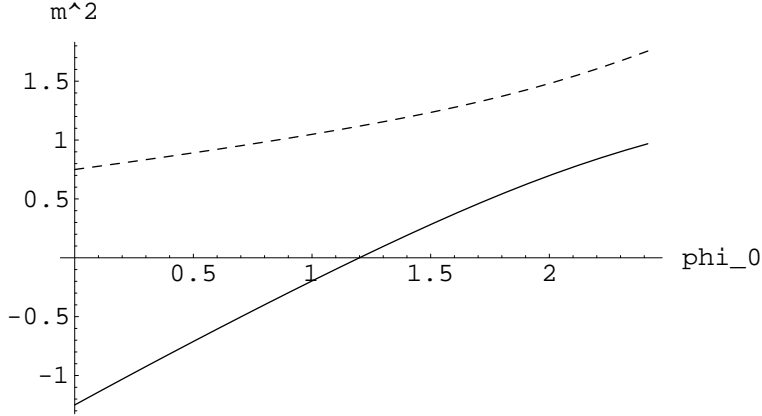}
\end{center}
\caption[]{\small The solid curve shows the flow
of the mass squared for the lowest mass state on
the lump. The flow begins at  $m^2=-5/4$. 
The dashed line shows
the flow of the massive state.} \label{f4}
\end{figure}

Consideration of the next level of 
approximation shows this is not a very precise result.
We  expand the level (2,4) potential around the
(2,4) expectation values given in \refb{24exp} to find
the quadratic form  
\be
V^{(2,4)}  \simeq -1.192 + 0.661 \phi_0^2 +  0.392 \phi_0 \phi_1
+ 0.702 \phi_1^2 + {\cal O}(\phi^3)\,.
\ee
After diagonalization one obtains two masses: $m^2 = 0.969, 1.756$.
Finally, we can use the exact expectation values for the fields,
as given in \refb{vacvalues} and insert them into the level (2,6)
potential finding the quadratic form 
\be
V^{(2,6)}  \simeq -1.183 + 0.671 \phi_0^2 +  0.413 \phi_0 \phi_1
+ 0.805 \phi_1^2 + {\cal O}(\phi^3) \,.
\ee
In this case  after diagonalization we find the following eigenvalues
and associated wave-functions (recall $\phi_0$ and $\phi_1$ are
associated to $\xi_0(x)$ and $\xi_1(x)$):
\ben
\label{mafter}
m_1^2 \simeq 1.04\,,  &&\hskip-10pt  0.81 \xi_0 -0.59 \xi_1 \,,\cr
m_1^2 \simeq 1.91\,,  &&\hskip-10pt  0.59 \xi_0 +0.81 \xi_1 \,. 
\een
As we can see, the wavefunction of the first state 
is mostly $\xi_0$ and that of the second state is mostly
$\xi_1$. We therefore expect the first state to be the
endpoint of the flow of the tachyon, 
and the second
state to be the endpoint of the flow of the massive state. Indeed,
this way the two mass levels do not cross under the flow.
In order to confirm this we have studied the flow of the masses
by letting $\phi_0$ vary, using the $\phi_1$ equation of motion
to fix $\phi_1$ as a function of $\phi_0$, and diagonalizing
the mass matrix for every value of $\phi_0 \in [0, \op_0]$.  The result
is shown in Fig.~\ref{f4}.   

Let us try to understand the meaning of what we have just
calculated. The rearrangement of the lump degrees of freedom
into the vacuum degrees of freedom  is based on the simple 
statement that the plane waves of the vacuum can be expanded 
in terms of the lump wavefunctions as:
\be
\label{expand}
e^{ipx} = \phi_0 (p) \,\xi_0(x)  + \phi_1 (p)\,\xi_1(x)  +
\psi_0(p) \,\eta_0(x) + \int dk\, \Phi_k(p)\,  \Xi_k (x)\,.
\ee
In here, the expansion coefficients $\phi_0(p), \phi_1 (p)$ and $\Phi_k(p)$
are readily calculable.  As we let the lump begin to flow into the vacuum, we
expect the potential to define a mass matrix that ceases to be diagonal
and whose mass eigenstates become linear combinations of the lump 
wavefunctions. As we approach the vacuum, those linear combinations must
approach the ones given in the above right hand side, for the pure
exponentials are the  eigenstates of the mass matrix around the vacuum. In 
fact, $m^2 = 1+ p^2$ for the wave $\exp(ipx)$.

It follows readily form \refb{expand} that
\be
\phi_0(p) = \sqrt{15\over 2}\,  {\pi (p^2 + {1\over 4})\over \cosh p\pi}\,,
\quad \phi_1(p) = 
\sqrt{3\over 2}\, \, {5\pi (p^2 - {3\over 20})\over \cosh p\pi}\,, 
\ee
giving a ratio of 
\be
\label{getp}
{\phi_1(p)\over \phi_0(p)} = \sqrt{5}\,\,
 {p^2 - {3\over 20}\over p^2 + {1\over 4}}\,\,.
\ee
We can now use the above result to test our flow. The first
state of \refb{mafter} has $m^2 = 1.04$ and $\phi_1/\phi_0 =
-0.59/0.81= -0.727$. Back in \refb{getp} this gives $p^2= 0.051$ and a mass
squared
$m^2 = 1 + p^2 \simeq 1.05$. This is in very good agreement with
the value $1.04$ found directly by diagonalization.
For the other state, we have $m^2 = 1.91$ and $\phi_1/\phi_0 = 1.37$.
Back in \refb{getp} this gives $p^2 \simeq 0.79$, and therefore
predicts  $m^2=1.79$, in reasonable agreement with the value $1.91$
obtained by diagonalization.

Since  momentum
dependence plays no role in the field theory potential,
as opposed to the case in string theory, the whole flow of
masses is due to mass terms changing. In string theory
both $m^2$ and $p^2$ terms change under the flow, and the resulting
flow of masses (determined from zeroes in the inverse propagator
$p^2 + m^2$) will be a combined effect of these two changes. 
Certainly mass squared terms alone can do the job in the field
theory case.

\sectiono{Large marginal deformations: moving the lump}
\label{s5}

In a recent work of Sen and the author \cite{0007153}
a puzzle was found. In order to study large marginal
deformations the authors calculated in the level
expansion the effective potential for the marginal
parameter. The marginal parameter was taken to be
the expectation value of the zero mode of the gauge
field on a D-brane, a Wilson line.  In the T-dual picture, this
marginal parameter is the parameter translating
the brane, and this viewpoint was emphasized in
the early analysis of \cite{0001201}.  Surprisingly, 
it was found in
\cite{0007153} that there is a critical value of 
the string field marginal parameter beyond
which its effective potential fails to exist. This
raised a puzzle. In conformal theory both
the Wilson line parameter
or the translation parameter take values from
zero to infinity. How does the string field theory
manage to describe the physics with a finite range
marginal parameter~? Two options were discussed: the
string field critical value corresponds to 
(i) infinite CFT parameter, or (ii) finite CFT
parameter. The first possibility was considered
to be better, the second worse, for it seemed
to imply that SFT could not describe 
in a single patch all of CFT moduli space. 

The interpretation here, to be substantiated below, is
that indeed the critical value corresponds to a 
finite CFT parameter (possibility (ii)) but 
this {\it does not} mean that SFT does not cover
CFT moduli space. What happens is that there
is a double valued relation: 
to every SFT parameter we associate two CFT 
parameters. As the CFT parameter 
parameter grows from zero to infinity the SFT parameter 
grows from zero to a maximum value and then decreases
back to zero! Of course, as the SFT marginal parameter starts
to decrease fields higher in the level expansion 
must take larger expectation values.

To substantiate the above claim let us consider
the  lump field theory, where the field $\psi_0$
associated with the wavefunction $\eta_0(x)$ plays
the role of a marginal parameter. Since $\eta_0$
is proportional to the derivative of the profile,
$\psi_0$ is the parameter associated with translating
the lump along the $x$ coordinate. Thus, $\psi_0$
plays the role of the string field marginal parameter
in the present model.

We begin by considering the potential \refb{discpot}
restricted
to the tachyon and the marginal parameter:
\be
\label{margpot}
V(\phi_0, \psi_0) = - {5\over 8} \, \phi_0^2 
 + {175\over 8192} \sqrt{15\over 2}\, \pi\, \phi_0^3 
+  {75\over 2048} \sqrt{15\over 2}\, \pi\,  \phi_0  \psi_0^2\,. 
\ee
In order to find the effective potential for the marginal
field we must use the tachyon equation to find $\phi_0 (\psi_0)$:
\be
\label{twosols}
\phi_0^\pm (\psi_0) = -{1\over 210\,{\sqrt{30}}\,\pi } \left( -4096 \pm
{\sqrt{16777216 - 
          756000\,{\pi }^2\,{\psi_0}^2}} \,\right) 
\ee
which indeed shows that the domain of definition
of the effective potential for $\psi_0$ cannot exceed 
\be
|\psi_0 |\leq {512\over 15\,\pi }\sqrt{2\over 105} \simeq 1.4995\,.
\ee
This is exactly parallel to the situation in string field theory
\cite{0007153}.  There are two branches to the solution. In the
notation of \cite{0007153}, the marginal branch is that where
the tachyon begins with zero expectation value and corresponds to
the top sign in \refb{twosols}. In the stable branch, corresponding to the
bottom sign, the tachyon begins with its expectation value for the
stable vacuum.  
The effective potential on the marginal branch is obtained substituting
$\phi_0^+$ into \refb{margpot}. Expanding the result for small $\psi_0$
we find 
\be
V_{eff}^M(\psi_0) = {16875\,{\pi }^2\over {4194304}}\, {\psi_0}^4 + {\cal
O}{(\psi_0)}^6\,.
\ee
As in \cite{0001201,0007153} this effective potential has a leading
quartic term, but would be expected to be identically zero in the 
exact solution.  For the stable branch we find: 
\be
V_{eff}^S(\psi_0) = -\frac{1048576}{99225\,{\pi }^2} + 
  \frac{5\,{\psi_0}^2}{7} + 
  {{\cal O}(\psi_0)}^4 \,,
\ee
and we see that to this first rough approximation,
we get  $m^2(\psi_0) = 10/7 \simeq 1.429$ for
the marginal mode around the stable vacuum.  Including the
effects of the field $\phi_1$ changes the above results
moderately. Working to level $(2,4)$ in these fields and 
exactly on $\psi_0$ we find that the domain of definition
of the marginal effective potential shrinks down to
about $|\psi_0 |\leq  1.426$. Substituting the exact
values $\op_0$ and $\op_1$ at the vacuum into \refb{discpot} we can read
a mass term for $\psi_0$ of  
\be
\label{tfap}
m^2(\psi_0) = {5535 \pi^2\over 32768} \simeq 1.667\,.
\ee
In fact, the exact version of this result is readily
obtained. We are simply studying the mass term associated
to the fluctuation $\tilde \phi = \psi_0 \,\eta_0(x)$ around the stable
vacuum  $\phi=1$.  The contribution to (minus) the action from this
fluctuation is quadratic and given simply by  ${1\over 2} \int dx  
((\tilde \phi')^2 + (\tilde \phi)^2)$.  Numerical integration gives
\be
m^2(\psi_0) = 1.714286\,,
\ee
for the mass of this mode around the stable vacuum. Note that 
the approximation with two fields in \refb{tfap} was  good.
In string field theory a nonvanishing mass squared was
found for the fluctuation mode of the marginal parameter 
around the stable. This was consistent with the expectation
that the vacuum has no marginal deformations \cite{0007153}
and with the expectation that the gauge field on the brane
dissappears after condensation \cite{0008033}.

\begin{figure}[!ht]
\leavevmode
\begin{center}
\epsfbox{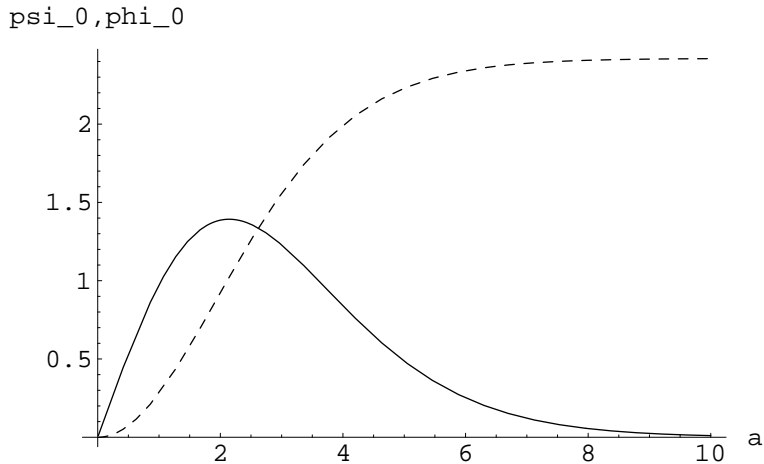}
\end{center}
\caption[]{\small The solid curve shows the expectation
value of the lump marginal parameter $\psi_0$ as 
a function of the displacement $a$ of the lump. Note
that the marginal parameter first increases, reaches
a maximum, and then decreases. 
The dashed line shows the expectation value of
the tachyon field $\phi_0$ as the lump is displaced.
Note that as the displacement is large the expectation
value of $\phi_0$ reaches the critical value $\op_0$
associated to the stable vacuum.} \label{f5}
\end{figure}

We now turn to an explanation for the finite range
of marginal parameters. As we have seen above, it happens
in the field theory model for the ``marginal'' state
$\psi_0$, so it is not a strictly stringy phenomenon. 
Consider the deformation that moves the lump a distance
$a$ to the left. From the viewpoint of the lump, this 
is a field fluctuation of the form $\op (x+a) - \op(x)$
since added to the lump profile $\op(x)$ it gives us 
$\op (x+a)$, representing the lump at $x=-a$. This fluctuation
is therefore to be expanded as usual:
\be
\label{expshift}
\op (x+a) - \op(x)  =  \phi_0 (a) \,\xi_0 (x) + \psi_0 (a) \,\eta_0 (x) +
\cdots\,,
\ee 
where the expansion coefficients are $a$-dependent. For the marginal
mode one finds
\be
\label{intres}
\psi_0(a) = \int_{-\infty}^{\infty} dx \Bigl(\op (x+a) - \op(x)\Bigr) \,
\eta_0 (x)  = \int_{-\infty}^{\infty} dx \,\op (x+a) \, \eta_0(x) \,,
\ee 
where we used the $x\to -x$ symmetry of $\op (x)$ and antisymmetry of
$\eta_0(x)$.  This integral is readily done for small $a$. To this 
end first use \refb{lumpprofile} and \refb{summdisc} to obtain
\be
{d\op\over dx} = \sqrt{6\over 5} \,\eta_0 (x)\,,
\ee
which gives the precise normalization relating the derivative
of the profile to the massless mode representing the marginal
operator.  With this result, the integral \refb{intres} is
readily done for small $a$ giving:
\be
\psi_0(a)  = \sqrt{6\over 5} \,\,a + {\cal O} (a^2) \,.
\ee
This is the expected linear relation between the
``SFT marginal parameter'' $\psi_0$ and the ``CFT
marginal parameter'' $a$. On the other hand it is 
manifest from the integral expression \refb{intres} for $\psi_0(a)$
and the fact that $\eta_0(x)$ is localized in $x$, that
for sufficiently large $a$ the overlap between the lump,
now centered at $a$ and $\eta_0(x)$ will go to zero. Thus, very
large $a$ will correspond to small $\psi_0(a)$. This is the
double valued relation we mentioned before. In Fig.~\ref{f5} we
show, in continuous line, the value of $\psi(a)$. Indeed, the plot
begins straight but $\psi_0$ reaches a maximum value of
$\psi_0^{max} \simeq 1.39$ for $a\simeq 2.3$.  This maximum value
is in good agreement with the level (2,4) approximation that resulted
in $|\psi_0 |\leq  1.426$. For values of $a$ larger than $a=2.3$
the magnitude of $\psi_0$ decreases.  

The expansion in \refb{expshift} suggests what is happening
when $a > 2.3$.  In this case, some higher level
fields, indicated by the dots will acquire larger expectation
values. Thus for each value of $\psi_0$ we obtain two values
of $a$, the small one realized with small expectation values
for the high level fields, and the large one realized with large 
expectation values for some high level fields. 

We now investigate the expectation value of the tachyon
field $\phi_0(a)$ as a function of the displacement. In this
case it follows from \refb{expshift} that
\be
\label{inttach}
\phi_0(a) = \int_{-\infty}^{\infty} dx \Bigl(\op (x+a) - \op(x)\Bigr) \,
\xi_0 (x) \,.
\ee 
For small $a$ the terms in parenthesis are well approximated by $a$ times
the derivative of the profile. Since $\xi_0$ is even, the integral
cancels, and we must have $\phi_0(a) \sim a^2 + {\cal O}(a^4)$.
On the other hand for large $a$ the first term in the parenthesis
becomes irrelevant to the integral, and we simply get the overlap
of the tachyon wavefunction with the profile. This is precisely
given by $\op_0$  (recall \refb{exactcond}).  Thus  
$\phi_0(a) \to  \op_0$ as $a \to \infty$.  This behavior is shown
in Fig.~\ref{f5}, where the $\phi_0(a)$ is shown as a dashed line.
The intuition behind this result is clear, by the time the lump
is far away we  have recovered the vacuum in
the vicinity of $x=0$. This requires giving the tachyon the
expectation value $\op_0$.

\begin{figure}[!ht]
\leavevmode
\begin{center}
\epsfbox{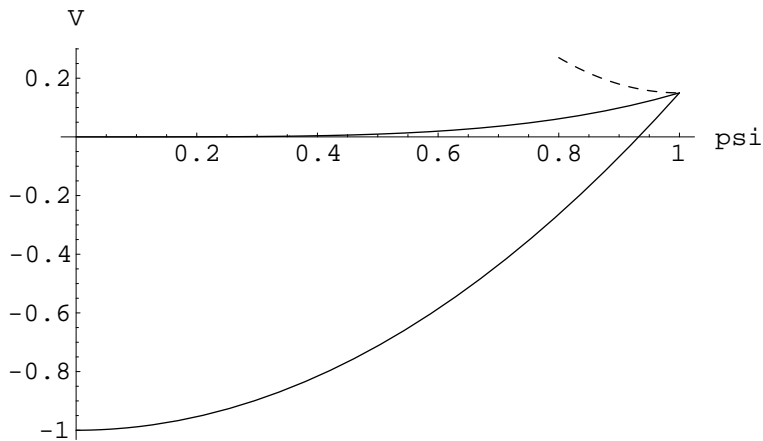}
\end{center}
\caption[]{\small A sketch of various physical branches
as the normalized marginal parameter $\psi$ varies from
zero to its maximal value. The vertical axis denotes the
potential energy. The solid curve
hugging the real line represents the marginal branch. It
represents moving the lump all the way to $a\simeq 2.3$. 
The bottom curve (solid) represents the stable branch.
The dashed curve, represents a missing
branch describing 
displacements of the lump larger than $a=2.3$.} \label{f6}
\end{figure}

We can now interpret more fully the branch structure found
in \cite{0007153}, as exemplified by Fig.~1 of that work. 
For convenience we have included Fig.~\ref{f6}
with a sketch of the branches. We show the potential
energy $V$, normalized  to the energy of the lump, and
$\psi_0$ normalized to its maximum value.   
The solid curves
were discussed in \cite{0007153}, and here we have 
added a dashed curve. The solid curve closest
to the real line represents the marginal branch. 
In perfect approximation it should be flat. When the 
maximal value of $\psi_0$ is attained it represents
the lump displaced to some finite distance ($a=2.3$ in the
model). The bottom solid curve is the stable branch, at its
leftmost point it intersects $V=-1$ and represents the vacuum.
In the present model it is clear what that curve represents.
As we go from the branch point downhill to the left we are
simply letting the lump, centered at $a=2.3$ decay away. 
The dashed line represents the branch where the lump is
moved beyond $a=2.3$. In this branch we expect  the
vev's of higher level fields to increase. This branch
has not been identified yet in the string field theory,
but the study of the field theory model suggests very
strongly that it will be found there.

\sectiono{Conclusions and open questions}

We considered $\phi^3$ field theory and
used the exact lump solution and
the complete solvability of the fluctuation problem to 
produce a detailed model where the decay of an
unstable lump is seen from the viewpoint of the
field theory living on the lump itself.

There are many analogies with string field theory
studies of D-brane annihilation.
In particular, the multiscalar potential on the
lump is nontrivial, and the vacuum state can be
explored in the level expansion.   Indeed, if
did not have the exact expression for the tachyonic
condensate in the model, we would have been 
hard pressed to find it from the explicit form of
the cubic multiscalar potential. The problem becomes 
simple only with the realization that the multiscalar 
condensate is basically the expansion of the lump 
profile in terms of the lump wavefunctions. 
It may be possible to apply these lessons 
fruitfully to the study of the 
tachyon condensate in string field theory.

One of the deepest questions with regards to
string field theory is whether or not the string
field is big enough.  Indeed, before tachyon
condensation and brane annihilation was shown to be 
described correctly by string field theory, it was
thought that the string field would not reach far
enough to describe this non-perturbative vacuum.
Now we know this is not the case.  In the recent work
of Sen and the author \cite{0007153} the possibility
resurfaced that maybe the string field is not
big enough, as large marginal deformations 
could possibly be beyond reach of the string field.  
Based on similar circumstances in the toy model, 
we have been able to propose a natural resolution
where a different branch of the solution space
contains the large marginal deformations. It thus
seems likely that the string field, after all,
is big enough.

\bigskip

\noindent {\bf Acknowledgments}:
I am indebted to J. Goldstone for instructing me on
reflectionless potentials and on their relevance to 
field theory solitons. I thank R. L. Jaffe for his
insights on the role of tachyons in the annihilation
of solitons and for many useful discussions. 
I am very grateful to J. Minahan for his interest in this work and 
for many stimulating conversations and detailed discussions.
Comments and suggestions by A. Sen are also gratefully
acknowledged. Finally, I thank R. Jackiw for help identifying
references on the early literature. This work was
supported in part by DOE contract \#DE-FC02-94ER40818.

\end{document}